
\texsis
\twelvepoint
\paper
\Eurostyletrue
\def\yo1{{{F_\pi}^2}}
\def\fpi2{{{F_\pi}^4}}
\def\laf1{{\int{ds}'g(s')\phi_{\gamma}(s')}}
\def\alfs{{\alpha_{0}}}
\def\alfp{{\alpha_{1}}}
\def\alfse{{\alpha_{2}}}
\def\lam1{{\int_0^\infty}}
\def\me{{ {m_\epsilon}^2}}
\def\mr{{ {m_\rho}^2}}
\def\mt{{ {m_T}^2}}
\def\lom{{\log({-s\over{\mu^2}})}}
\def\loma{{\log({-s\over{\mu_i^2}})}}
\def\lomp{{\log({s\over{\mu^2}})}}

\def\leb{{\biggl\lbrack}}
\def\les{{\biggl\lbrace}}
\def\rib{{\biggr\rbrack}}
\def\ris{{\biggl\rbrace}}
\def\tD{{\tilde D_{2}}}
\def\bta{{\beta_1}}
\def\btb{{\beta_2}}
\referencelist
\reference{chan1} For a review, see M. Chanowitz, \journal Ann. Rev. Nuc. Sci;
                  38,323 (1988)
\endreference
\reference{*chan1a} For more recent results, see G.L. Kane, preprint
  UM-TH-92-23, (1992)
\endreference
\reference{mms}  See, for example, B.R. Martin, D. Martin, and G. Shaw,
            Pion-Pion Interactions in Particle Physics (Academic, London, 1976)
            and references therein
\endreference
\reference{dr} See, for example, A. Dobado, M. Herrero, and T.N. Truong,
          \journal Phys. Lett. B;235,129 (1990)
\endreference
\reference{*dra} D.A. Dicus and W.W. Repko, \journal Phys.
                 Rev. D;42,3660 (1990), and references therein
\endreference
\reference{sw2}  S. Weinberg, \journal Phys. Rev. Lett.;65,1177 (1990)
\endreference
\reference{*sw2a} T.N. Pham, preprint, CNRS-91
\endreference
\reference{bc} S.R. Beane and C.B. Chiu, in preparation
\endreference
\reference{csb}  C.B. Chiu, E.C.G. Sudarshan, and G. Bhamathi, \journal Phys.
                Rev. D;45,884 (1992)
\endreference
\reference{bg1}  L.S. Brown and R.L. Goble, \journal Phys. Rev. Lett.;20,346
               (1968)
\endreference
\reference{bg2}  L.S. Brown and R.L. Goble, \journal Phys. Rev. D;4,723
               (1971)
\endreference
\reference{*bg2a} For an Higgs sector analysis see
                  R. Rosenfeld \journal Phys. Rev. D;42,126 (1989)
\endreference
\reference{chan2} M. Chanowitz, preprint LBL-32938 (1992)
\endreference
\reference{bps}  S.N. Biswas, T. Pradhan and E.C.G. Sudarshan, \journal Nucl.
                Phys. B;50,269 (1972)
\endreference
\reference{sw}  S. Weinberg, \journal Physica; 96A, 327 (1979)
\endreference
\reference{leh} H. Lehmann, \journal Phys. Lett. B;41,529 (1972)
\endreference
\reference{sf} S.C. Frautschi, Regge Poles and S-matrix Theory
                  (Benjamin, New York, 1963) p. 13
\endreference
\reference{adler}  S. Adler, \journal Phys. Rev.;140,B736 (1965)
\endreference
\reference{yy} Y. Yamaguchi, \journal Phys. Rev.;95,1628 (1954)
\endreference
\reference{*yya} Y. Yamaguchi and Y. Yamaguchi, \journal \it ibid.;95,1635
                 (1954)
\endreference
\reference{vs}  V. Srinivasan et al., \journal Phys. Rev. D;12,681 (1975)
\endreference
\reference{nnb}  N.N. Biswas et al., \journal Phys. Rev. Lett.;47,1378 (1981)
\endreference
\reference{*nnba} N.M. Cason et al., \journal Phys. Rev. D;28,1586 (1983)
\endreference
\reference{md} M. David et al., unpublished
\endreference
\reference{*mda} G Villett et al., unpublished
\endreference
\reference{bh}  B. Hyams et al., \journal Nucl. Phys. B;64,134 (1973)
\endreference
\reference{*bha} W. Ochs, Thesis, University of Munich (1973)
\endreference
\reference{ber}  V. Bernard, N. Kaiser, and U. MeiBner, \journal Nucl. Phys.
               B;364,283 (1991)
\endreference
\endreferencelist
\titlepage
\title
Direct-Channel Dominance in Goldstone Boson Scattering in the Resonance Region
\endtitle
\author
S.~R.~Beane\Footnote*{Internet: sbeane@ccwf.cc.utexas.edu} and C.~B.~Chiu
Center for Particle Physics and Department of Physics
The University of Texas at Austin, Austin, TX 78712
\endauthor
\abstract
We consider the scattering of Goldstone bosons in the range of intermediate
energies, and in particular focus our attention on the scattering of pions in
the $\rho$-resonance region. The chiral perturbation series is obtained
to order $E^4$ from a chiral $SU(2)\times SU(2)$ invariant effective
Lagrangian. At order $E^4$, we isolate
the low-energy manifestation of heavy particle exchange in the
s-channel. We find that the contributions of this type to the I=1 and I=2
amplitudes have a one parameter dependence.
This observation provides a symmetry rationale for Chanowitz's
recent observation that in a chiral model with an explicitly coupled $\rho$,
the I=1 and I=2 channels are strongly correlated.
In order to realize, in a
quantitative way, the fact that in the resonance region the direct-channel
dominates, we make use of a simple and intuitive unitarization scheme.
This allows us to derive a renormalization constant-independent relation
for the I=1 and I=2 phase shifts. With the assumption of a
$\rho$-resonance, this relation determines the position of
a pole at euclidean momentum in the I=2 channel. Through an analysis based on
the theory of redundant poles and the Adler sum rule we show that
this ``tachyon'' pole actually represents a physical
contribution to the unitary amplitude which accounts for $\rho$ and $\epsilon$
exchange in the u-channel.
\endabstract
\endtitlepage
\vfill\eject                                     
\section{Introduction}
\superrefsfalse
In recent times, there has been a great deal of interest in the
scenario of a strongly interacting Higgs sector\ref{chan1}. The
breakdown of
all model-independent predictions in the energy region where new
physics is expected to appear has led to a proliferation of speculations.
On one front, the familiar isomorphism between the pions and the longitudinal
components
of the standard model gauge bosons has led many physicists to make use
of methods that were in fashion long ago.
Unitarization and completeness sum rules were widely used in the past, in the
context of hadronic physics, with varying degrees of success\ref{mms}.
These approaches usually involved relating the physics at threshold, where
chiral perturbation theory is valid, to
physics in the range of intermediate energies where little is known for sure.
These two methods
have also been applied to the standard model Higgs sector\ref{dr}\ref{sw2}.
In this paper we will consider a synthesis of model-independent results
obtained from chiral perturbation theory and old fashioned methods in a way
that we believe is quite new. Since we have quantitative
statements to make, we will, for obvious reasons, concentrate on pion
interactions. We wish to stress that it is not our intention to improve
or replace chiral perturbation theory. Rather, we are interested in determining
whether any new insight into the intermediate energy regime can be gained by
merging effective field theory techniques with technology that was a staple
of the S-matrix program.

If one calculates to a given order in chiral perturbation theory with
goldstone bosons alone, one obtains an amplitude whose properties include
manifest crossing symmetry and perturbative unitarity.
Of course, in the presence of resonances, the direct-channel dominates in the
energy region where experiments are performed. Therefore, an interesting
question to address is whether the crossing properties of the undetermined
parameters that appear in chiral perturbation theory reveal any interesting
substructure. This paper addresses this issue. It is clear that
in order to extract quantitative statements from this program, we will
necessarily have to make use of a model which trades manifest crossing
symmetry for exact unitarity. Some would argue
that this type of approach makes use of unitarity to yield meaningful
information, in flat contradiction with current lore.
We argue that this point is more subtle than one would expect.
In a separate work\ref{bc}, we argue that the ``theorem'' (or rule of
thumb), which states that
axiomatic properties like unitarity do not uniquely determine any
S-matrix element, provides a powerful constraint on the form that a
unitarized amplitude can take. It is interesting that $N\over D$ schemes
satisfy this constraint in a natural way, relegating any predictive power
to the neglect of classes of crossed diagrams which under certain
conditions can be assumed negligible.

One way of generating amplitudes of $N\over D$ form is by solving a
Schr\"odinger equation for a relativistic Hamiltonian (RH)\ref{csb}.
In the RH model that we use, the
effective potential consists of all direct-channel two-particle-irreducible
diagrams in the chiral expansion.
At the level of chiral perturbation theory there is exact crossing
symmetry, and as a result, the lowest order diagrams associated with resonance
exchange
in the direct- and crossed-channels appear at the same order in the power
counting. On the other hand,
in the RH model, the crossed-channel contributions enter as a
perturbation on the direct-channel. This is clearly a result of
breaking crossing symmetry in favor of unitarity. We refer to this rather
obvious property as ``direct-channel dominance''. One result of direct-channel
dominance is that, in a
certain sense, the RH amplitudes with the lowest order effective potentials
can be thought of as modified low-energy theorems since these amplitudes
offer the possibility of
revealing substructure that is hidden when crossing symmetry is restored in
an approximate way. That is, by restoring crossing symmetry,
we necessarily introduce undetermined parameters.
Therefore, a quantitative statement
extracted from this model is not strictly a consequence of enforcing
unitarity, but
rather a result of assuming that contributions to the scattering amplitude
arising from resonance exchanges in the crossed-channel are small.

This point of view provides insight
into the meaning of unitarization and is supported by $\pi$-$\pi$ phase shift
phenomenology, where it has long been known
that simple bubble-sum approximations which contain no left-hand cut
contributions yield solid agreement with the available data. In fact,
the amplitudes that we obtain are similar to a
parametrization derived from an effective range expansion about the current
algebra low-energy theorems\ref{bg1}. In an interesting study of this
parametrization, it was found that
minimizing the amount by which crossing symmetry is
violated leads to an approximate degeneracy in the pole
structure of the
I=1 and I=2 amplitudes\ref{bg2}. We will show that this degeneracy is encoded
in the undetermined coefficients of chiral perturbation theory.

We find certain
relations among the renormalization constants in the  RH amplitudes by
requiring that the expanded unitary amplitudes agree with the chiral
perturbation series to order $s^2$. Although the
parameters that appear in chiral perturbation theory at this order are
undetermined, the
relations among these parameters as they appear in the states of definite
isospin provide non-trivial information. In particular, we find that the RH
phase shifts satisfy the following renormalization constant independent
relation:

$$\kappa_{21}\equiv
s(\cot\delta_{2}(s)-\cot\delta_{1}(s))={{-128\pi\yo1}}+O(s^2). \EQN in1$$

This relation implies a strong correlation between the I=1 and I=2
amplitudes. In a recent work\ref{chan2}, Chanowitz has shown that in
a unitarized chiral model with an explicitly coupled $\rho$
(or techni-$\rho$), the
I=1 and I=2 channels are strongly correlated.
We show that this ``complementarity''
is a direct consequence of constraints imposed by chiral symmetry near
threshold.

If we assume the existence of a $\rho$-resonance in the I=1 channel, we find
that
the I=2 RH amplitude has a negative-s pole on the physical sheet, which
would seem to correspond to an unphysical tachyon pole.
We make use of a mechanism proposed by Biswas, Pradhan and Sudarshan
(BPS)\ref{bps} in order to reinterpret the tachyon pole.
If one replaces the original theory with a new theory, wherein one introduces
the negative-s pole in the effective potential, the new theory yields the
same
amplitude as the original theory, except that the negative-s pole no longer
appears as a zero of the denominator function. Instead, it is to be identified
as an effective pole representing left-hand cut contributions.$^1$\vfootnote1{
In potential theory such a pole is said to be redundant
since it doesn't contribute to the completeness relation.
See S.T. Ma, Phys. Rev. 69 (1964) 668.} As an
independent test of this hypothesis, we consider the possible
effective pole contribution to the Adler sum rule. We find that the sum rule
requires that the position of the I=2 effective
pole be of the order of the $\rho$ mass. This result is consistent with the
BPS interpretation.

The plan of the remainder of the paper is as follows. In sect. 2, we
obtain the chiral perturbation series to order $E^4$ by way of a chiral
Lagrangian. In sect. 3, we recall
the relativistic Hamiltonian model with separable potentials, and construct
the scattering amplitudes with tree level effective potentials. We then make
use of the BPS mechanism in sect. 4. In sect. 5, we consider the
implications of the Adler sum rule. Finally,
in sect. 6  we give a brief summary and conclusion. Our $\pi$-$\pi$ conventions
and the details of the relativistic Hamiltonian model are relegated to
appendices.
\section{Chiral Perturbation Theory}
In the chiral limit, the $SU_L(2)\times SU_R(2)$ invariant Lagrangian,
including terms with four derivatives, is given by

$${\cal L}={\cal L}^{(2)}+{\cal L}^{(4)}, \EQN chi1 $$
with

$$
{\cal L}^{(2)}={\yo1 \over 4}{\Tr(\partial_\mu\Sigma\partial^\mu\Sigma^
  \dagger)},     \EQN chi2a$$
and

$$ {\cal L}^{(4)}={C_1\Tr(\partial_\mu\Sigma\partial^\mu\Sigma^\dagger)
                       \Tr(\partial_\nu\Sigma\partial^\nu\Sigma^\dagger)}
                 +{C_2\Tr(\partial_\mu\Sigma\partial_\nu\Sigma^\dagger)
                       \Tr(\partial^\mu\Sigma\partial^\nu\Sigma^\dagger)}.
                 \EQN chi2b $$
The Goldstone boson fields are contained within the field variable

$${\Sigma}={exp(i \vec\tau\cdot\vec\pi \over F_\pi)},
                 \EQN chi3 $$
which has the transformation property

$$\Sigma\rightarrow{g_L \Sigma g_R\dagger}\quad ;\quad g_{L,R}\in SU_{L,R}(2),
\EQN chi3a $$
under $SU_L(2)\times SU_R(2)$.
$C_1$ and $C_2$ are undetermined constants which characterize the
underlying theory at low energies. The amplitude to order $E^4$ may
be obtained from tree and 1-loop graphs using  ${\cal L}^{(2)}$ and tree
graphs using
${\cal L}^{(4)}$\ref{sw}. A straightforward Feynman diagram calculation yields

$$
\eqalign{
{A(s,t,u)} &={s \over \yo1}+{4\over \fpi2}{\lbrack 2C_1(\mu^2)s^2
                      +C_2(\mu^2)(t^2+u^2)\rbrack}
                   +{1 \over{(4\pi)^2\fpi2}}\bigl\lbrack{-s^2\over2}\
                  log({-s\over{\mu^2}})\cr
           & -{1\over 12}(3t^2+u^2-s^2)\log({-t\over{\mu^2}})
           -{1\over 12}(3u^2+t^2-s^2)\log({-u\over{\mu^2}})\bigr\rbrack .\cr}
\EQN chi4 $$
In order to more clearly display the crossing properties of the undetermined
parameters, we rewrite the $E^4$ term in the form originally given by
Lehmann\ref{leh},

$$
\eqalign{
&{A^{(4)}(s,t,u)}={-1 \over{6(4\pi)^2\fpi2}}\biggl\lbrack{3s^2}\
                  \lbrack\log({-s\over{\mu^2}})+\beta_1(\mu^2)\rbrack\cr
& +t(t-u)\lbrack\log({-t\over{\mu^2}})
+\beta_2(\mu^2)\rbrack
+u(u-t)\lbrack\log({-u\over{\mu^2}})
+\beta_2(\mu^2)\rbrack\biggr\rbrack .\cr}     \EQN chi4a $$
Here we see that $\beta_1$ is related to heavy particle exchange in the
s-channel whereas $\beta_2$ is related to heavy particle exchange in the
t- and u-channels.
With the conventions given in the Appendix, we project
out the partial wave amplitudes of definite isospin. We find,

$$\EQNalign{
{a_0}\equiv{a_{00}(s)}=&{\alpha_{0}s}\les 1-{{\alpha_{0}s}\over{\pi}}
                   \leb\log({-s\over{\mu^2}})+{1\over4}(3\bta+\btb)\rib \cr
                         &-{{\alpha_{0}s}\over{\pi}}
                         \leb{7\over18}\log({s\over{\mu^2}})-{11\over108}
                         +{1\over18}(3\bta+4\btb)\rib\ris ,
\EQN chi5;a \cr
{a_1}\equiv{a_{11}(s)}=&{\alpha_{1}s}\les 1-{{\alpha_{1}s}\over{\pi}}
                         \leb\log({-s\over{\mu^2}})+\btb\rib \cr
                         &+{{\alpha_{1}s}\over{\pi}}
                         \leb\log({s\over{\mu^2}})-{1\over3}
                         +(3\bta-2\btb)\rib\ris ,
\EQN chi5;b \cr
{a_2}\equiv{a_{20}(s)}=&{\alpha_{2}s}\les 1-{{\alpha_{2}s}\over{\pi}}
                         \leb\log({-s\over{\mu^2}})+\btb\rib \cr
                         &-{{\alpha_{2}s}\over{\pi}}
                         \leb{11\over9}\log({s\over{\mu^2}})-{25\over54}
                         +{1\over 9}(6\bta+5\btb)\rib\ris
\EQN chi5;c \cr   }
$$
where

$$\alfs\equiv{1\over{16\pi\yo1}},\quad
  \alfp\equiv{1\over{96\pi\yo1}},\quad{\rm and}\quad
  \alfse\equiv {-1\over{32\pi\yo1}}. \EQN chi6 $$
Each curly bracket consists of three terms, corresponding to the low-energy
theorem, and the loop contributions in the direct- and the
crossed-channel respectively. {\it Note that we have been careful to preserve
the crossing properties of the terms in T$^{(4)}_{I}$(s,t,u) with
undetermined coefficients}.

The key observation of this paper is that the contributions associated with
direct-channel loop diagrams in the I=1 and I=2 amplitudes depend on the same
renormalization constant. In the spirit of direct-channel dominance, we expect
that, if the $E^4$ term in the chiral expansion is at all representative of
a general trend, then
there should be a strong correlation between the
I=1 and I=2 amplitudes in the energy region where perturbative unitarity
breaks down. This is consistent with Chanowitz's complementarity.
In order to make quantitative statements to this effect, we
make use of the RH model.
\section{The Relativistic Hamiltonian Model}

In our formalism (the details are relegated to an Appendix), the effective
potential is given by all direct-channel
two-particle irreducible diagrams in the chiral expansion. In accord with
direct-channel dominance, we consider the tree-level effective potential,
given by the low-energy theorem,

$$ {-\pi}V_i(s)={\alpha_is}. \EQN rh1 $$
The corresponding RH amplitude is given by

$$ t_i={{-\pi V_i}\over{1-I_i}}, \EQN rh2 $$
with

$$I_i(s)=-<V_i>={{\alpha_i}\over{\pi}}<s>, \EQN rh3 $$
where

$$<f(s)>\equiv \int{{f(s')ds'}\over{s'-s-i\epsilon}}. \EQN rh3a$$
The dispersion integral $<s>$, through an appropriate regularization
procedure\ref{csb},
defines an analytic function. $I_i(s)$ takes the form:

$$-{{\alpha_i}\over{\pi}}s\leb\lom+R_i\rib+b_i. \EQN rh4 $$
We choose $b_i=0$ as our definition of the renormalized pion decay constant.
We then find that

$$t_{i}(s)={{\alpha_i s}\over{1+{{\alpha_i s}\over\pi}[\lom+
     R_{i}(\mu^2)]}}={1\over{ {1\over{\alpha_i s}}+{1\over\pi}(\lomp+R_i)-i}}.
 \EQN rh5 $$
In terms of the phase shifts,

$$\cot\delta_i={1\over{\alpha_i s}}+{1\over\pi}(\lomp+R_i).
\EQN rh7 $$
In order to determine the $R_i$, we expand the denominator of the RH
amplitude. We then match the form of the direct-channel piece of the fourth
order amplitude, obtained from the chiral Lagrangian, with the form of the
truncated RH amplitude. Inspection of \Eq{chi5} yields,

$$\EQNalign{
R_{0}(\mu^2)&={1\over4}(3\bta{(\mu^2)}+\btb{(\mu^2)}),\,{\rm and}\EQN rh8;a \cr
     R_{2}(\mu^2)&=R_{1}(\mu^2)=\btb{(\mu^2)}.
      \EQN rh8;b \cr }$$
The novel feature of this unitarization scheme is the one
parameter dependence of the I=1 and I=2 amplitudes. In terms of phase shifts
we obtain the difference,

$$\kappa_{21}\equiv s(\cot\delta_{2}(s)-\cot\delta_{1}(s))={{-128\pi\yo1}},
\EQN rh9$$
a result independent of $\bta$ and $\btb$. In order to display the accuracy
of this result, we write

$$\kappa_{21}={{-128\pi\yo1}}+O(s^2), \EQN rh9a$$
where within our formalism, the terms of order $s^2$ constitute a
perturbation due to crossed-channel contributions.$^2$\vfootnote2{ For the
sake of comparison, we note that the [1,1] Pad\'e approximant yields

$\kappa_{21}={{-128\pi\yo1}}+P(s;\bta,\btb)$
where

$P(s;\bta,\btb)=s\leb{-1\over\pi}({20\over9}\lomp -{43\over54})+
{1\over\pi}(\btb-{2\over3}\bta)\rib =O(s).$
The K-matrix method also yields an O(s) remainder. See ref.\ref{bc} for
a critique of the Pad\'e method.}

At this point we must say something about the spectrum of resonances that
couple to the $\pi$-$\pi$ system. We assume the existence of a
$\rho$-meson in the I=1 p-wave channel. This assumption fixes the I=2
amplitude to depend only on the $\rho$ mass. We also adopt the conventional
assumption that there exists an $\epsilon$-meson in the I=0 s-wave channel
(Whether the $\epsilon$ serves as a convenient parametrization of strong
final state interactions among pions or is the $\it veritable$ thing is of
no concern to us here.)
We choose our renormalization point such
that the resonance occurs "near" $s=\mu^2$ (i.e. when the narrow width
approximation is valid). Defining

$$D_{i}(s)\equiv {1+{{\alpha_i s}\over\pi}[\lom +
     R_{i}(\mu^2)]}, \EQN h1$$
we assume that at $s=\mu^2$, $ReD_{i}(s)=0$, which gives

$$R_{i}(\mu_i^2)={-\pi \over {\alpha_{i}\mu_i^2}}, \EQN h2$$
and

$$D_{i}(s)=1-{s\over {\mu_i^2}}+{{\alpha_i s}\over\pi}\loma .
  \EQN h3$$
Finally, we obtain

$$D_{0}(s)=1-{s\over \me}+{\alpha_0s\over\pi}
            \log({-s\over \me});  \EQN h4$$

$$D_{1}(s)=1-{s\over \mr}+{\alpha_1s\over\pi}
            \log({-s\over \mr});  \EQN h5$$

$$D_{2}(s)=1-{(-3)s\over \mr}+{\alpha_2s\over\pi}
            \log({-s\over \mr}).  \EQN h6$$
The factor of 3 in \Eq{h6} is a consequence of the fact that
${{\alpha_2}\over{\alpha_1}}=-3$. The nearby zero of $D_{2}$ may be obtained
numerically. We find the position of this pole to be

$$s=-{m_T}^2=-(0.31)\mr,\quad{\rm or}\quad m_T\simeq(0.6)m_\rho. \EQN h7$$
A zero of the denominator which occurs at negative values of s corresponds to
a ``tachyon'' pole. This would appear to be a physically unacceptable solution.
\section{The BPS Mechanism [10]}
By way of the crossing matrix given in the Appendix, we find

$$T(I_s=2)={1\over3}T(I_u=0)+{1\over2}T(I_u=1)+{1\over6}T(I_u=2). \EQN h8$$
This isospin decompostion reveals that the I=2 S-wave amplitude
should be dominated by a left-hand cut contribution associated with the
exchange of $\rho$ and $\epsilon$.
However, recall that within the RH formalism crossed-channel resonance
exchange contributions arise from the parameters that appear at order
$E^4$ in the chiral expansion. Nevertheless,
we can explicitly include left-hand cut contributions in the potential by way
of an effective pole\ref{sf}. In particular,
consider a new effective potential for the I=2 channel with a pole at the
tachyon position,

$$\tilde V_{2}(s)={{-\alpha_{2}s\mt}\over{\pi(\mt+s)}}
                  \equiv {{-\kappa s}\over{\mt+s}}. \EQN h9$$
The denominator function of the new theory, subtracted at $s=-\mt$ is given by

$$
\eqalign{
\tD (s)-\tD (-\mt)&=-\kappa\lam1{{ds's'}\over{\mt+s'}}
                    \leb{1\over{s'-s}}-{1\over{s'+\mt}}\rib \cr
                  &=-\kappa (s+\mt)\lam1{{ds's'}\over{(\mt+s')^2(s'-s)}} \cr
                  &=-\kappa (s+\mt)\lam1{{ds'}\over{(\mt+s')}}
                    \leb{1\over{(s'+\mt)}}+{s\over{(\mt+s')(s'-s)}}\rib. \cr}
\EQN h10$$
 Evaluating the integrals, we find

$$\tD (s)=\tD (-\mt)-\kappa+{{\kappa s}\over{(\mt+s)}}\log({-s\over\mt}).
\EQN h11$$
We choose our subtraction condition such that as $s\rightarrow 0$,
$\tilde t_{2}\rightarrow -\pi\tilde V(s)=-\pi V(s)$, or $\tD (0)=1$. Thus,

$$\tD (s)=1+{{\kappa s}\over{(\mt+s)}}\log({-s\over\mt})
         ={\mt\over{(s+\mt)}}\leb 1+{s\over\mt}+
          {{\alfse s}\over\pi}\log({-s\over\mt})\rib. \EQN h12$$
This leads to

$$\tilde t_{2}={{-\pi\tilde V_{2}}\over\tD}
              ={{\alfse s}\over{1+{s\over\mt}+{\alfse s\over\pi}
               \log({-s\over\mt})}}
              =t_{2}.
\EQN h13$$

Let us recollect what we have accomplished. We began with an
effective potential obtained from the leading term in chiral perturbation
theory. In order to remove the unphysical tachyon pole from the denominator
function,
we considered an alternative theory whose potential contains an
effective pole at $s$=$-\mt$, and therefore higher orders in an expansion in
powers of ${s\over\mt}$. With this new potential, we arrive at the same
amplitude that we had originally obtained. However, a physical
interpretation now exists.
It is well known that only the zeros of the D-function contribute to the
direct-channel spectrum. In the new theory, we see that the effective pole
naturally accounts for crossed-channel contributions.
In light of this reinterpretation, we can now legitimately proceed and
consider sum rule constraints. Of course, the sum rules necessarily
provide a consistency check of this reinterpretation.
\section{A Further Look at the I=2 Amplitude}
We recall that assuming the leading Regge trajectory for I=1
t-channel exchange has $\alpha_1(0)<1$ leads to an unsubtracted dispersion
relation for the I=1 t-channel amplitude. Evaluating at threshold
yields the Adler sum rule\ref{adler},

$${T^{(-)}\over s}\vert_{s=t=0}={1\over\yo1}={1\over\pi}\lam1 {ds\over s^2}
         Im\,\leb {2\over3}T_s^{(0)}+T_s^{(1)}-{5\over3}T_s^{(2)}\rib,
   \EQN f1$$
where in the notation of the Appendix,

$$T^{(-)}\equiv {1\over2}\leb{A(s,t,u)-A(u,t,s)}\rib={1\over2}\leb
{2\over3}T_s^{(0)}+T_s^{(1)}-{5\over3}T_s^{(2)}\rib, \EQN f2$$
and $T_s^{(i)}\equiv T(I_s=i)$. Rearrangement of \Eq{f1} yields

$$
{-1\over{2\yo1}}={1\over\pi}\lam1{ds\over s^2}Im\,\leb T_s^{(2)}\rib -
                    {1\over\pi}\lam1{ds\over s^2}
Im\,\leb{1\over3}T_s^{(0)}+{1\over2}T_s^{(1)}+{1\over6}T_s^{(2)}\rib \EQN f3$$

$$
\qquad\quad    ={1\over\pi}\lam1{ds\over s^2}Im\,\leb T_s^{(2)}\rib -
                    {1\over\pi}\lam1{du\over u^2}
Im\,\leb{1\over3}T_u^{(0)}+{1\over2}T_u^{(1)}+{1\over6}T_u^{(2)}\rib, \EQN f4$$
where in the last step we have used the relation

$$\lam1{ds\over s^2}Im\,\leb T(I_s=i)\rib=\lam1{du\over u^2}Im\,\leb
  T(I_u=i)\rib, \EQN f5$$
which states that the physics in the s-channel for $I_s=i$ is identical to
the physics in the u-channel for $I_u=i$. Using the crossing matrix given in
the Appendix, \Eq{f4} becomes

$${-1\over{2\yo1}}={1\over\pi}\lam1{ds\over s^2}Im\,\leb T(I_s=2)\rib -
                   {1\over\pi}\lam1{du\over u^2}Im\,\leb T(I_s=2)\rib .
\EQN f6$$
In an analogous manner we find,

$${1\over\yo1}={1\over\pi}\lam1{ds\over s^2}Im\,\leb T(I_s=0)\rib -
                   {1\over\pi}\lam1{du\over u^2}Im\,\leb T(I_s=0)\rib ,
\EQN f7$$
and

$${1\over{2\yo1}}={1\over\pi}\lam1{ds\over s^2}Im\,\leb T(I_s=1)\rib -
                   {1\over\pi}\lam1{du\over u^2}Im\,\leb T(I_s=1)\rib .
\EQN f8$$

The sum rules given by \Eq{f6}, \Eq{f7}, and \Eq{f8} are assumed to be valid
for the ``physical'' amplitude. Now we want to consider the implications of
these sum rules for the RH model amplitudes. We make use of the resonance
saturation approximation, so each integral is finite.
If we neglect distant tachyon contributions, the $I_s$=0 and $I_s$=1 RH
amplitudes contain no left-hand cut contributions. Therefore, \Eq{f5}, \Eq{f7},
and \Eq{f8} imply that

$${1\over\pi}\lam1{du\over u^2}Im\,\leb T(I_u=0)\rib\equiv
                   {1\over\pi}\lam1{ds\over s^2}Im\,\leb T(I_s=0)\rib
    ={1\over\yo1} \EQN f9$$
and

$${1\over\pi}\lam1{du\over u^2}Im\,\leb T(I_u=1)\rib ={1\over{2\yo1}}.
 \EQN f10$$
Using \Eq{f9}, \Eq{f10}, and \Eq{f4}, we obtain

$${-{1\over{2\yo1}}}=I_{ex}-({1\over{3\yo1}}+{1\over{4\yo1}}+{1\over6}I_{ex}),
\EQN f11$$
where $I_{ex}$ is the ``exotic spectral contribution'', i.e.

$$I_{ex}={1\over\pi}\lam1{ds\over s^2}Im\,\leb T(I_s=2;s)\rib =
         {1\over\pi}\lam1{du\over u^2}Im\,\leb T(I_u=2;u)\rib
         ={1\over{10\yo1}}. \EQN f12$$
We have evaluated the exotic spectral contribution in an indirect manner; that
is, by using the Adler sum rule together with the $I_s$=0 and $I_s$=1 RH
amplitudes. Next we evaluate the exotic contribution directly from the $I_s$=2
RH amplitude. The analytic structure of the $I_s$=2 RH amplitude is given
in fig. 1. By converting the right-hand cut integral into a clockwise contour
integration along C, we find

$$I_C\equiv I_{ex}={1\over{2\pi i}}\oint_C
ds\,{{T(I_s=2;s)}\over{s^2}}=I_o+I_A,
\EQN f13$$
where we have used the relation,

$$C_\infty =-C+C_o+C_A. \EQN f14$$
We now relax our renormalization condition of sect. 4. With

$$T(I_s=2,s)={{16\pi\alfse s}\over{1+{s\over\mt}+{\alfse s\over\pi}
               \log({-s\over\mt})}}, \EQN f15$$
we obtain

$$I_{ex}={16\pi\alfse}\leb 1-{1\over{\mt({1\over\mt}+{\alfse\over\pi})}}\rib
         \equiv{\zeta\over{2\yo1}}, \EQN f16$$
with

$$\zeta={{-{{\alfse\mt}\over\pi}}\over{ {{\alfse\mt}\over\pi}-1}},\quad or\quad
        \mt={-{\pi\over\alfse}}\cdot{\zeta\over{1+\zeta}}.\EQN f17$$
Using the values $F_{\pi}$=93 MeV, and $m_{\rho}$=770 Mev, we find that

$$\mt\approx 4.6{\zeta\over{1+\zeta}}\mr. \EQN f18$$

\Eq{f11} reveals the relative importance of the three isospin amplitudes.
The I=2 amplitude contributes ${1\over{12\yo1}}$, while
the I=0 and I=1 amplitudes contribute $-{7\over{12\yo1}}$.
That is, the exotic contribution
is only ${1\over7}$ of the non-exotic contribution and therefore, as
expected, the former is relatively insignificant.
Furthermore,
if we require that the sum rule given by \Eq{f12} be satisfied, we find that
$\zeta =0.2$, or from \Eq{f18}, $m_{T}\simeq (0.9)m_{\rho}$. This result is
consistent with our renormalization procedure result,
$m_{T}\simeq (0.6)m_{\rho}$ (see \Eq{h7}.) The fact
that the Adler sum rule requires the magnitude of the effective pole position
to be of the order of the $\rho$ mass provides strong support
in relating the tachyon which appears in the original RH model I=2
amplitude, to the exchange of $\rho$ and $\epsilon$ in
the u-channel.$^3$\vfootnote3{ Note that $\rho$, $\epsilon$ degeneracy follows
from assuming the KSRF relation for the
$\rho$ width together with a superconvergence relation for the I=2 t-channel
amplitude, as discussed in ref.\ref{mms}. See also ref.\ref{sw2}.}

As mentioned earlier, Brown and Goble\ref{bg2} found a similar
degeneracy in the pole structure of the I=1 and I=2 amplitudes by
minimizing the amount by which crossing symmetry is violated. They speculated
that this degeneracy could be the result of some underlying symmetry.
Of course, at that
time, it was not clear that unambiguous statements could be made beyond
tree-level in the chiral expansion. By considering the relations between the
renormalization constants that appear at order $E^4$ in the
chiral expansion, we have shown that the underlying symmetry resposible for
this degeneracy is in fact chiral symmetry, as realized in the range of
intermediate energies where the direct-channel dominates.

In fig. 2(a) and fig. 2(b) respectively, we display the I=0 and I=1 phase
shifts in the special case where $m_\epsilon$=$m_\rho$.
In fig. 3 we
display the I=2 phase shift for the two values of $m_{T}$ given above.  Our
model predictions are in agreement with the trend of the data.
\section{Summary and Conclusion}
We have shown that when contributions to the chiral perturbation series at
order $E^4$ are separated according to their crossing properties,
an interesting result follows. The
contributions associated with direct-channel heavy particle exchange in the
I=1 and
I=2 amplitudes have a one parameter dependence. In the resonance region, where
perturbative unitarity breaks down, the direct-channel dominates, implying a
strong correlation between the I=1 and I=2 channels. This provides a
symmetry argument for Chanowitz's notion of complementarity\ref{chan2}. We
explicitly illustrated
this mechanism by considering a relativistic Hamiltonian model with effective
potentials given by the lowest order amplitudes in chiral perturbation theory.

With the assumption of a $\rho$ resonance, the relation that we obtain for the
I=1 and I=2 amplitudes implies the existence of a ``tachyon'' pole in the I=2
amplitude whose ``absolute'' position is nearly degenerate with the $\rho$.
By way of the BPS
mechanism we showed that this negative-s pole
can be reinterpreted to be a pole of the effective potential which represents
$\epsilon$ and $\rho$ exchange in the crossed-channel.
In order to test this hypothesis, we considered the implications of
the Adler sum rule, which is derived using the ``physical'' amplitude,
and therefore incorporates crossing symmetry constraints. We found that the
Adler sum rule supports our hypothesis.
This result suggests that the degeneracy found by Brown and Goble\ref{bg2}
in the pole structure of the I=1 and I=2 amplitudes is a direct result of
constraints imposed by chiral symmetry near threshold.
Notwithstanding the neglect of all contributions to the effective
potential associated with heavy particle exchange in the crossed-channel,
chiral symmetry would appear to require the I=2 amplitude to ``mock up''
these contributions by way of an effective pole.
\showsectIDfalse
\section{Acknowledgements}
We thank Prof. E.C.G. Sudarshan for invaluable discussions. This work was
supported by the Department of Energy Grant No. DEFG05-85ER40200.
\showsectIDfalse
\appendix{A}{Appendix A$\colon$ The $\pi$-$\pi$ System}

The $\pi$-$\pi$ scattering amplitudes are determined by crossing symmetry in
terms of a single analytic function $A(s,t,u)$ as

$$T_{\alpha \beta ;\gamma \delta}={A(s,t,u)\delta_{\alpha \beta}
   \delta_{\gamma \delta}+A(t,s,u)\delta_{\alpha \gamma}\delta_{\beta \delta}+
   A(u,t,s)\delta_{\alpha \delta}\delta_{\beta \gamma}} \EQN A1$$
where

$$\EQNalign{ s&=(p_\alpha+p_\beta)^2, \EQN A3;a \cr
             t&=(p_\alpha-p_\gamma)^2, \EQN A3;b \cr
             u&=(p_\alpha-p_\delta)^2, \EQN A3;c \cr}$$
are the usual Mandelstam variables. These amplitudes can be decomposed
into amplitudes of definite isospin by using the standard techniques of
projection operators. We find,

$$\EQNalign{ T_0(s,t,u)&=3A(s,t,u)+A(t,s,u)+A(u,t,s), \EQN A4;a \cr
             T_1(s,t,u)&=A(t,s,u)-A(u,t,s), \EQN A5;b \cr
             T_2(s,t,u)&=A(t,s,u)+A(u,t,s), \EQN A6;c \cr}$$
where the crossing matrix that relates the amplitudes in the s-channel
to those in the u-channel is given by

$$\left(\matrix{T_0^s\cr T_1^s\cr T_2^s\cr}\right)=
  \left(\matrix{{1\over3}&{-1}&{5\over3}\cr
                {-{1\over3}}&{1\over2}&{5\over6}\cr
                {1\over3}&{1\over2}&{1\over6}\cr}\right)
  \left(\matrix{T_0^u\cr T_1^u\cr T_2^u\cr}\right).
  \EQN A6a$$
In turn, the partial waves can be projected out using,

$$a_{IJ}(s)={1\over{32\pi}}{\int_{-1}^1}d(\cos\theta)P_J(\cos\theta)
            T_I(s,t,u), \EQN A7$$
where for elastic scattering (in the chiral limit),

$$a_{IJ}=e^{i\delta_{IJ}}\sin\delta_{IJ}. \EQN A8$$
The total cross section follows from the optical theorem,

$$\sigma_{IJ}^{tot}(s)={{16\pi}\over s}Im(a_{IJ}(s)) \EQN A8$$
and the phase shifts are obtained from

$$\cot\delta_{IJ}(s)={{Re(a_{IJ}(s))}\over {Im(a_{IJ}(s))}}. \EQN A9$$
The narrow width approximation for the total cross section is given by

$$\sigma_R^{tot}(s)={{16\pi^2m_R}\over s}(2J_R+1)\Gamma_R\delta(s-m_R^2).
  \EQN A10$$

\showsectIDfalse
\appendix{B}{Appendix B$\colon$ The Separable Potential Model [6]}
We use the c.m. energy square variable, s, the orbital angular
momentum, $\ell$, and
isospin, i to label the $\pi$-$\pi$ amplitudes. Since the Hamiltonian
matrix is
diagonal in $\ell$ and i, we can study each channel separately. For each
channel, the Hamiltonian takes the form:

$$H=H_0+H_I. \EQN sep1$$
The free Hamiltonian term is given by

$$H(s,s')=s\delta(s-s'). \EQN sep2$$
We assume the separable potential form\ref{yy}

$$H_I(s,s')=\mp g(s)g(s'), \EQN sep3$$
where without loss of generality, $g(s)$ can be taken to be real. The
effective potential is given by

$$V(s)=H_I(s,s)=\mp g^2(s). \EQN sep4$$
The minus and plus signs correspond to attractive and repulsive potentials
respectively. The scattering amplitude is given by\ref{yy}

$$T(s)={{-\pi V(s)}\over{\beta (s)}}, \EQN sep5$$
where

$$\beta=1+\int{{V(s')ds'}\over{s'-s-i\epsilon}}. \EQN sep6$$
It is clear that the solution, \Eq{sep5}, is a simple bubble-sum with a
kernel given by V(s).
\section{Figure Captions}
Fig.(1)
Analytic structure of I=2 S-wave R.H. amplitude and complex plane contours
for the I=2 sum rule.

Fig.(2)
(a) I=0 S-wave phase shift from R.H. model amplitude with tree level effective
potential compared with experimental data of ref.\ref{vs}(points),
ref.\ref{nnb}(open circles), and ref.\ref{md}(open squares).
(b) I=1 P-wave phase shift from R.H. model amplitude with tree level effective
potential compared with experimental data of ref.\ref{vs}(points),
ref.\ref{nnb}(circles), and ref.\ref{bh}(crosses). All data points are
taken from ref.\ref{ber}.

Fig.(3)
I=2 S-wave phase shifts from R.H. model amplitude with tree level effective
potential compared with experimental data of ref.\ref{vs}(points) and
ref.\ref{nnb}(open circles). The dotted phase shift follows from
$m_{T}\simeq (0.6)m_{\rho}$, and the solid phase shift follows from
$m_{T}\simeq (0.9)m_{\rho}$. All data points are
taken from ref.\ref{ber}.
\nosechead{References}
\ListReferences
\end